\documentstyle[preprint,aps,prc]{revtex}
\newcommand{\nuc}[2]{\mbox{$^{#1}$#2}}
\begin{document}
\draft
\title{Dilepton production from virtual bremsstrahlung
 induced by proton capture}
\author{D. Van Neck, A.E.L. Dieperink and O. Scholten}
\address{Kernfysisch Versneller Instituut, Zernikelaan 25,
9747 AA Groningen, The Netherlands}
\maketitle
\begin{abstract}
Dilepton production following radiative capture of a proton on a
nuclear target is studied in the Impulse Approximation. The cross
section is decomposed in terms of four time-like nuclear structure
functions through a longitudinal-transverse separation of the nuclear
current. Using a simple PWIA model, cross sections and conversion
factors are calculated for capture reactions $p+n\rightarrow d+e^+ e^-$,
 $p+p\rightarrow \nuc{2}{He}+e^+ e^-$
and $p+ \nuc{11}{B}\rightarrow \nuc{12}{C}+e^+ e^-$
at proton energies of 100-200 MeV. The result is compared with
a recent measurement of the conversion factor for $p+ \nuc{11}{B}$.
\end{abstract}

\section{Introduction}

The detection of dilepton pairs has been shown to be a powerful
technique in heavy-ion reactions. In heavy-ion reactions at
higher energies there are several sources of dileptons, (i)
nucleon-nucleon bremsstrahlung, (ii) Dalitz decay of resonances and
mesons, (iii) meson annihilation. The interest here lies in the study
of medium effects and possible phase transitions.
 In order to separate the contributions from  the first and the latter
 two mechanisms, it is important to be able to describe the cross section
for the bremsstrahlung accurately. This can be done best for simple
systems like $p+A$ or $p+p(n)$.

Recently it was shown \cite{wies} that detection of dilepton pairs
even at relatively low energies ($E$ = 100 MeV) is feasible, e.g. in
the radiative capture of protons on $^{11}$B. Interestingly it was found
 that the con\-version ratio
$R=\sigma(\mbox{dilepton})/\sigma(\mbox{photon})$
(at fixed photon angle) varied with excitation energy in the final
nucleus.

In ad\-dition to being of inter\-est for the general under\-standing
of the
vir\-tu\-al brems\-strahlung process in the nuclear medium
the investigation of production of dileptons in $p+A$ is of interest for
specific reasons. For example a new aspect of dilepton production
(compared to real photon production) is the presence of a virtual
longitudinally
polarized photon. Furthermore this process contains information
on the time-like nucleon form factor which is not easy to obtain
otherwise.

To compute the cross section for dilepton production at lower energies
(below the pion production threshold)
it is extremely useful
to decompose the amplitudes in longitudinal (L) and transverse
(T) components
which have a different dependence on the kinematic observables,
analogously to the $(e,e'p)$ reaction which can be viewed as the
spacelike counterpart. As a result one can
(in the impulse approximation) make a separation of the cross section
into transverse and longitudinal parts.

The interest in isolating the longitudinal part stems from the fact
that it is assumed to correspond to a simple charge operator
with a monopole
(C0) component which is expected to be insensitive to exchange currents.
In fact the study of the C0 component is also of interest in related
fields of physics, e.g.\ in muon catalyzed fusion \cite{fus}\newline
$\nuc{3}{H}+p \rightarrow ^{4}\mbox{He}^*
\rightarrow ^{4}\mbox{He} + e^+ + e^-$.

In this paper we compute the cross section for dilepton production
for the special limit of radiative capture of a proton on a nuclear
target
in the PWIA (Plane Wave Impulse approximation).
We decompose the cross section in terms of
longitudinal, transverse
and interference terms. Then we express the cross-section
ratio of virtual to real photons in terms of L and T conversion factors.
Using a simple one-body prescription for the nucleon current
we consider radiative capture for $p+n\rightarrow d+e^+ e^-$,
$p+n\rightarrow \nuc{2}{He}+e^+ e^-$ and
$p+ \nuc{11}{B}\rightarrow \nuc{12}{C}+e^+ e^-$.

\section{Derivation of the cross section}

We consider the reaction $p+A\rightarrow (A+1) +e^+ +e^-$ (with $A$
an unspecified nuclear system) in the Impulse Approximation (IA).
The four-momenta involved (see figure \protect\ref{fig1})
will be denoted
by $p^{\mu}$ (incoming proton), $k^{\mu}_A ,k^{\mu}_{A+1}$ (target and
residual nuclear system), $q^{\mu}_1 ,q^{\mu}_2$ (outgoing electron and
positron). The
exchanged virtual (time-like) photon has  4-momentum
$Q^{\mu}=q^{\mu}_1 +q^{\mu}_2$
 and  an invariant mass $M^2 =Q\cdot Q=Q^{2}_0 -Q^2 >0$.

The squared amplitude for this process in the one-photon approximation
is written as
\begin{equation}
|{\cal T}|^2 = \frac{e^4}{(Q\cdot Q)^2}|j\cdot J|^2=\frac{e^4}{M^4}
J^{*}_{\mu}J_{\nu}L^{\mu\nu},
\end{equation}
with $j^{\mu}$ the lepton and $J^{\mu}$ the nuclear current.
After summation over the $e^+ e^-$ polarizations, the lepton tensor
is equal to (see e.g. \cite{itz}):
\begin{eqnarray}
\label{e2}
L^{\mu\nu}&=&\frac{1}{4m_e^2}\mbox{Trace}\left(
(q\!\!\!/_2 -m_e )\gamma^{\mu}(q\!\!\!/_1 +m_e)\gamma^{\nu}\right)\\
&=&\frac{1}{m_e ^2}\left(
q^{\mu}_1 q^{\nu}_2 +q^{\mu}_2 q^{\nu}_1 - g^{\mu\nu}
(q_1 \cdot q_2 +m_e^2)\right)\nonumber\\
&=&\frac{1}{2m_e ^2}\left(Q^{\mu}Q^{\nu}-q^{\mu}q^{\nu}-M^2 g^{\mu\nu}
\right),\nonumber
\end{eqnarray}
with $q^{\mu}=q^{\mu}_1 -q^{\mu}_2$ the relative $e^+ e^-$ momentum.
The appearance of a negative energy projection operator
$(m_e -q\!\!\!/_2 )$ in eq. (\ref{e2})
is the only difference with the corresponding expression in $(e,e')$
scattering.

Taking current conservation ($J\cdot Q=0$) into account, the invariant
amplitude can be expressed as:
\begin{equation}
\label{e4}
|{\cal T}|^2 = \frac{e^4}{M^4}\frac{1}{2m_e ^2}\left(
-|J\cdot q|^2 - M^2 J^{*}\cdot J\right).
\end{equation}

It is our aim to express the cross section for $(p,e^+ e^-)$,
 analogously to the well-known LT (longitudinal-transverse)
decomposition
for the $(e,e'p)$ process \cite{bof,fru}, in terms of four
independent structure functions.

We introduce the L and T components of the spatial part of the
nuclear current:
\begin{equation}
\vec{J}=\vec{J}_L +\vec{J}_T ,
\end{equation}
with
\begin{equation}
\vec{J}_L =\frac{(\vec{J}.\vec{Q})}{Q^2}\vec{Q}.
\end{equation}

Furthermore we  decompose $\vec{J}_T$ in two transverse directions:
$\vec{J}_T =J_{+1}\vec{e}^{\,*}_{+1} +J_{-1}\vec{e}^{\,*}_{-1}$, with
$\vec{e}_{\pm 1}=\mp \frac{1}{\sqrt{2}}(\vec{e}_x \pm i\vec{e}_y)$
defined in the reference system with the $z$-axis along $\vec{Q}$ and
the $y$-axis along $\vec{Q}\times\vec{p}$
(see figure \protect\ref{fig2}).

The LT separation of the terms in eq.\ (\ref{e4}) reads as:
\begin{eqnarray}
\label{lt1}
J^{*}\cdot J&=&-\frac{M^2}{Q_{0}^2}|J_L|^2 -|J_T|^2\\
J\cdot q&=&-\frac{M^2}{Q_{0}^2}q\cos \theta_q J_L
+\frac{1}{\sqrt{2}}q \sin \theta_q (J_{+1}e^{-i\varphi_q}
-J_{-1}e^{i\varphi_q}),
\end{eqnarray}
in terms of the solid angle $\Omega_q =(\theta_q,\varphi_q)$
of $\vec{q}$  in the
reference frame of figure \protect\ref{fig2}.
Note that we have used the continuity equation
\begin{equation}
\label{con}
Q J_L = Q_0 J_0
\end{equation}
to replace the charge density
$J_0$ by the longitudinal component $J_L$ of the nuclear current.
The reason for this will be explained in section \ref{s3}.

The differential $A(p,e^+ e^- )A+1$ cross section reads as:
\begin{equation}
\label{cs1}
d\sigma(p,e^+ e^-) = \frac{d\vec{q}_1 d\vec{q}_2 m^{2}_e}
{(2\pi )^6 (q_1)_0 (q_2)_0}
\frac{d\vec{k}_{A+1}}{(2\pi )^3 }\frac
{(2\pi )^4 \delta^{(4)}(k_{A+1}+q_1 +q_2 -p-k_A )}
{4\sqrt{(p\cdot k_A )^2 -m^2 m^{2}_A}}|{\cal T}|^2 .
\end{equation}
We will evaluate this cross section in the CM frame.
Introducing the invariant mass $M$ and momentum $\vec{Q}$ of the virtual
photon as new variables, the cross section (\ref{cs1})
can be written  as:
\begin{equation}
\label{cs2}
d\sigma(p,e^+ e^-)=\frac{1}{64\pi^2 s}\frac{Q}{p}
P_{e^+ e^-}|{\cal T}|^2 d\Omega dM^2 ,
\end{equation}
with $\sqrt{s}$ the total CM energy, $\Omega$ the solid angle of
$\vec{Q}$ in the CM frame, and $P_{e^+ e^-}$ the phase space
factor for the decay of the virtual photon into the $e^+ e^-$ pair.
The latter quantity is given by:
\begin{equation}
P_{e^+ e^-}=\int \frac{m_{e}^2 d\vec{q}_1 d\vec{q}_2
\delta^{(4)}(Q-q_1 -q_2)
}{(2\pi)^3 (q_1 )_0 (q_2 )_0}.
\end{equation}
It is convenient to evaluate $P_{e^+ e^-}$ in the rest frame
of the decaying virtual photon:
\begin{equation}
\label{16}
P_{e^+ e^-}=\frac{m_{e}^2} {(2\pi)^3}
\frac{u}{2} d\Omega'_q ,
\end{equation}
with $\Omega'_q =(\theta'_q ,\varphi'_q )$ the solid angle of the
relative $e^+ e^-$ momentum
in the rest frame of the virtual photon and $u=\sqrt{1-4m^{2}_e /M^2 }$.

The relation between $\Omega_q$ and $\Omega'_q$ can be found  by boosting
the relative $e^+ e^-$ momentum along $\vec{Q}$ with velocity $Q/Q_0$:
\begin{eqnarray}
\label{lt2}
\varphi'_q &=&\varphi_q ,\\
uM\cos \theta'_q &=&\frac{Q_0}{M}q\cos \theta_q -\frac{Q}{M}q_0
=\frac{qM}{Q_0}\cos \theta_q ,\nonumber\\
uM\sin \theta'_q &=&q\sin \theta_q .\nonumber
\end{eqnarray}
The angle $\varphi'_q$ represents the angle between the plane of
$(\vec{p},\vec{Q})$ and that of the dilepton pair.
The angle $\theta'_q$ is related to the asymmetry in the energies of the
$e^+ e^-$ pair, since
\begin{equation}
\cos \theta'_q =\frac{(q_1 )_0 -(q_2 )_0}{uQ},
\end{equation}
and plays a similar role as the electron scattering
angle in the $(e,e'p)$ reaction.

This becomes evident if we write the CM cross section after LT
decomposition, using eqs. (\ref{lt1}-\ref{lt2}) ($\alpha$ is the
fine-structure constant):
\begin{eqnarray}
\label{e12}
\frac{d\sigma(p,e^+ e^-)}{d\Omega dM d\Omega'_q}
&=&\frac{\alpha^2}{64\pi^2 s}\frac{Q}{p}
\frac{u}{M\pi}
\left\{ (1-u^2 \cos^2 \theta'_q )W_L \right.\nonumber\\
&&+(1-\frac{u^2}{2}\sin^2 \theta'_q )W_T \nonumber\\
&&+\frac{u^2}{2}\sin^2 \theta'_q \cos(2\varphi'_q )W_{TT}\nonumber\\
&&\left. +u^2\sqrt{2}\sin (2\theta'_q )\cos\varphi'_q W_{LT}\right\}.
\end{eqnarray}
The nuclear structure functions $W_i$ are given by:
\begin{eqnarray}
\label{struc}
W_L (p,Q,\theta,M)&=&\frac{M^2}{Q_0^{2}}|J_L|^2\\
W_T (p,Q,\theta,M)&=&|J_T|^2\nonumber\\
W_{TT}(p,Q,\theta,M)&=&
2\Re(J_{+1}J_{-1}^* ) \nonumber\\
W_{LT}(p,Q,\theta,M)&=&\frac{M}{Q_0}2\Re
(J_L J_{+1}^*  -J_L J_{-1}^* ).\nonumber
\end{eqnarray}
The different structure functions can be experimentally separated
through the dependence on the dilepton angles
$\theta'_q$ and $\varphi'_q$
of each term
in (\ref{e12}).

Based on eq. (\ref{e12}) one can derive various integrated
 cross sections. Integration over the out-of-plane angle $\varphi'_q$
of the $e^+ e^-$ pair makes the LT and TT interference terms (the last
two terms in eq. (\ref{e12}) )
vanish. A further integration over the asymmetry $\cos \theta'_q$
of the pair leads to
\begin{equation}
\label{ef1}
\frac{d\sigma(p,e^+ e^-)}{d\Omega dM} = \frac{\alpha^2}{64\pi^2 s}
\frac{Q}{p}
\frac{4}{M}u(1-\frac{u^2}{3})
\left(W_L +W_T \right).
\end{equation}

In many situations there is experimental information on the $(p,\gamma)$
 cross sections for real photons. Therefore it is of interest to
introduce
 the conversion factor $R(M,\theta)$, defined
  as the ratio of the cross sections for
emission of a virtual and a real photon in the same
direction $\Omega$:
\begin{equation}
\label{ef2}
R(M,\theta)=\frac{d\sigma(p,e^+ e^-)}{d\Omega dM} /
\frac{d\sigma(p,\gamma)}{d\Omega }
=R_L (M,\theta) + R_T (M,\theta)
\end{equation}

The CM cross section for the $A(p,\gamma)A+1$  process
in the denominator of eq. (\ref{ef2}) is given by
\begin{equation}
\label{e14}
\frac{d\sigma (p,\gamma)}{d\Omega}=\frac{\alpha}{64\pi^2 s}
\frac{k}{p} 4\pi|J_T |^2 ,
\end{equation}
with $\vec{k}$ the momentum of the real photon.
The longitudinal and transverse conversion factors become ($i$=L,T):
\begin{equation}
R_i (M,\theta)=\frac{Q}{k}\frac{\alpha}{\pi}u(1-\frac{u^2}{3})\frac{1}{M}
\frac
{W_i (p,Q,\theta,M)}
{W_T (p,k,\theta,0)}.
\end{equation}

In the following we will focus on properties of the conversion factors
$R_i$. In particular we will explore
its dependence on $M,\theta$ and nuclear structure.

\section{The nuclear current in PWIA}
\label{s3}

For this purpose we will use the non-relativistic
one-body current operator, which has the form:
\begin{eqnarray}
<\!\hat{J}_0 (\vec{Q},M)\!>&=&G_E (M^2)\\
<\,\hat{\!\!\vec{J}}(\vec{Q},M)\!>&=&\frac{1}{2m}\left(G_E (M^2)
(\vec{p}_i +\vec{p}_f)
+G_M (M^2) \vec{Q}\times i\vec{\sigma} \right),
\end{eqnarray}
taken between momentum eigenstates $\vec{p_i}$ and $\vec{p_f}$, with
$\vec{Q}=\vec{p_i}-\vec{p_f}$.

The nucleon mass is $m$
and $G_E ,G_M$ are the Sachs nucleon form factors in the time-like
region. For the relatively small values of $M$ ($M<100$ MeV)
that we will consider we use the continuation in the time-like
region of the dipole fit for the nuclear form factors. We neglect
the off-shellness of the captured proton.

The nuclear transition current in eq. (\ref{e4}) is then given by
\begin{equation}
\label{ntc}
\vec{J}(\vec{Q},M)
=<\!A+1|\,\,\hat{\!\!\vec{J}}(\vec{Q},M)|A,\vec{p}\,m_s \!>,
\end{equation}
with $|A>$ and $|A+1>$  non-relativistic wave functions of the
target and residual nucleus. Note that the use of non-relativistic
wave functions introduces
an extra normalization factor $(2k_{A}^0)(2k_{A+1}^0)(2p^0)$ in
our expressions
 (\ref{e12}-\ref{e14})
for the cross section.

In contrast to the conventional PWIA treatment of the $(e,e'p)$
reaction, where $J_L$ is eliminated,
we retain the longitudinal component $J_L$ of the nuclear current
instead of the charge density $J_0$. Both are in principle related via
the continuity equation (\ref{con}) if the initial and final state
are true eigenstates of the nuclear hamiltonian (and if the used
current and charge operators are consistent with the same hamiltonian).
 It is known \cite{can,ama,che}
that, due to the non-orthogonality of the bound-state
wave function and the plane-wave scattering state,
the PWIA treatment of the
transition charge density leads to incorrect results
in the region of small three-momentum transfer Q (even
for large proton energies),  since in PWIA $J_0
(Q\rightarrow 0) \neq 0$. In the $(p,e^+ e^-)$ reaction the longitudinal
response favours large invariant masses of the virtual photon, which
means small three-momentum transfer $Q=\sqrt{Q_0^2 - M^2}$.
 The use of $J_L$ at least guarantees the
correct small-Q, large-p behaviour of the longitudinal nuclear response,
as shown by the analysis of Amado et al.\ \cite{ama}.

\subsection{Factorized form of the current for medium and heavy nuclei}

If we make the assumption (similar to the IA treatment of the $(e,e'p)$
 reaction) that the incoming proton emits the photon, the
matrix element of the current between initial and final
nuclear states (eq.(\ref{ntc})) becomes (neglecting CM motion):
\begin{equation}
\vec{J}(\vec{Q},M)=\int \!d\vec{p_i }\int \!d\vec{p_f }\int \!d\sigma
\varphi^* (\vec{p_f },\sigma)
\delta(\vec{Q}-\vec{p_i}+\vec{p_f})<\,\hat{\!\!\vec{J}}(\vec{Q},M)\!>
\psi_{\vec{p} m_s}(\vec{p_i },\sigma),
\end{equation}
with
$\varphi$ the overlap function between $|A>$ and $|A+1>$
and $\psi_{\vec{p}}$ the wave function of the incident proton in
momentum space.

Taking a plane-wave description for the incident proton,
$\psi_{\vec{p}}$ reduces to a delta-function, and the current becomes:
\begin{equation}
\vec{J}(\vec{Q},M)=\frac{1}{2m}\int d\sigma \varphi^*
(\vec{p}-\vec{Q},\sigma)
\left( G_E (2\vec{p}-\vec{Q}) +G_M \vec{Q}\times i\vec{\sigma} \right)
\chi_{m_s}(\sigma).
\end{equation}

It can be shown, by a similar analysis as in the case of the
$(e,e'p)$ process \cite{fru}, that after averaging  over the spins of
the incident proton and target nucleus and summing over
the spin of the residual nucleus, the four structure
functions in PWIA have a common factor in which the dependence
on the
nuclear structure of the target and residual nucleus is contained.
The structure functions  in eq. (\ref{struc}) become,
after this spin summation:
\begin{eqnarray}
W_L  &=& \frac{1}{2(2J_A +1)}\frac{M^2}{Q_0^{2}}\frac{(2\pi)^3}{(2m)^2}
G_{E}^2 (2p\cos\theta -Q)^2
S(|\vec{p}-\vec{Q}|) \\
W_T  &=& \frac{1}{2(2J_A +1)}
\frac{(2\pi)^3}{(2m)^2}(G_{E}^2 4p^2 \sin^2 \theta + 2G_{M}^2
Q^2) S(|\vec{p}-\vec{Q}|) \\
W_{LT} &=& \frac{1}{2(2J_A +1)}
\frac{M}{Q_0}\frac{(2\pi)^3}{(2m)^2} G_{E}^2
(-4\sqrt{2})\, p \sin\theta
(2p\cos\theta -Q) S(|\vec{p}-\vec{Q}|) \\
W_{TT}&=&\frac{1}{2(2J_A +1)}
\frac{(2\pi)^3}{(2m)^2}G_{E}^2 (-4)p^2 \sin^2 \theta\,
 S(|\vec{p}-\vec{Q}|).
\end{eqnarray}
As a result, the total cross-section is also factorized into a
kinematical part and a nuclear structure part.

The common function $S$ is related to the spectral function for addition
of a proton to the target nucleus in its ground state.
Its explicit form reads as:
\begin{equation}
S(k)=\frac{1}{4\pi}\sum_{lj}|<\!J_{A+1}||c^{+}_{lj}(k)||J_A\!>|^2,
\end{equation}
with the reduced matrix element defined as in \cite{tal}.
If the ground state of the target can be described as a pure one-hole
state $h$ with respect to a spherical closed-shell residual nucleus,
we have
\begin{equation}
S(k)=\frac{2j_h +1}{4\pi}|\phi_h (k)|^{2},
\end{equation}
with the hole orbital $\phi_h (k)$ normalized as
\begin{equation}
\int dk k^2 |\phi_h (k)|^{2} = 1.
\end{equation}

\subsection{Nuclear current for the $A=2$ system}

For two-body systems, CM motion can of course be treated
exactly. We now also take photon emission by both nucleons into
account.

For the purpose of comparing $p+n$ and $p+p$ capture we first assume
a pure $l=0$ state  $\phi_0$ (no D-state admixture) as
the spatial part of the internal deuteron wavefunction.
For the quasi-bound $^{2}\mbox{He}$ system we take the same
spatial wavefunction
$\phi_0$,
but replace the spin triplet of the deuteron by a singlet state.
We used the $l=0$ part of the parametrization of the deuteron
wave function by Machleidt et al. (table 11 in \cite{mac}).

It is interesting to study
the effects of the antisymmetrization in case of identical nucleons.
If we couple the spins of the two-nucleon initial state to $S'$, and
denote by $S$ the spin of the final bound state, the nuclear current
 is given by
\begin{eqnarray}
\label{2bod}
J_L &=&\frac{1}{2m}\left(2p\cos \theta(G^{(1)}_E \phi_0^{(1)}-G^{(2)}_E
\phi_0^{(2)})
 - Q(G^{(1)}_E \phi_0^{(1)}+G^{(2)}_E
\phi_0^{(2)})\right)\delta_{SS'}\delta_{M_S M_{S'}}\nonumber\\
J_{\pm 1}&=&\mp \frac{1}{2m}\left(
\sqrt{2}p\sin \theta
(G^{(1)}_E \phi_0^{(1)}-G^{(2)}_E
\phi_0^{(2)})\delta_{SS'}\delta_{M_S M_{S'}}\right.\nonumber\\
&&\left.
-Q(1-\delta_{S0}\delta_{S'0})S_{\pm}\,
(G^{(1)}_M \phi_0^{(1)}+(-1)^{S+S'}G^{(2)}_M
\phi_0^{(2)})\right),
\end{eqnarray}
with $\phi^{(i)}_0=\phi_0(k_i)$ and
$\vec{k}_1 =\vec{p}-\vec{Q}/2$, $\vec{k}_2 =\vec{p}+\vec{Q}/2$.
 The spin factor $S_{\pm}$ is given by:
\begin{equation}
S_{\pm}(SM_S ,S'M_{S'})=
<\!{ \frac{1}{2}\,\frac{1}{2}}\,S'M_{S'}|\sigma_{\pm 1}(1)|
{ \frac{1}{2}\,\frac{1}{2}}\,SM_S\!>.
\end{equation}
It is clear from eq. (\ref{2bod}) that interference effects can occur
between the contributions from the two nucleons to the current.
This will
be further discussed in section \ref{s4.1}.

For the case of $p+n$ capture we will also
 give results with a more realistic treatment of the deuteron
wave function in which the D-state is included. The expressions for the
nuclear structure functions, as defined in eq. (\ref{struc}), then
become (after averaging over initial and summing over final
spin):
\begin{eqnarray}
\label{wl}
W_L &=&\frac{(2\pi)^3}{4}\left(
\frac{3}{4\pi}\sum_i \sum_l |C^{(i)}_l|^2 +\frac{3}{2\pi}
\sum_l C^{(1)}_l C^{(2)}_l P_l (\cos \omega_{12})\right)\\
W_T &=&\frac{(2\pi)^3}{4}\left(
\frac{3}{2\pi}\sum_i \sum_l (|E^{(i)}_l|^2
+ |M^{(i)}_l|^2)\right.\nonumber\\
&&+\frac{1}{\pi} \sum_l (3E^{(1)}_l E^{(2)}_l
+M^{(1)}_l M^{(2)}_l )P_l (\cos \omega_{12})\nonumber\\
&&-\frac{\sqrt{2}}{\pi}(M^{(1)}_0 M^{(2)}_2 P_2 (\cos \theta_{2})
+M^{(1)}_2 M^{(2)}_0 P_2 (\cos \theta_{1}))\nonumber\\
&&-\frac{2}{5}\sqrt{2}\sqrt{7}M^{(1)}_2 M^{(2)}_2
[Y_2 (\Omega_1)\otimes Y_2 (\Omega_2)]^2_0 \nonumber\\
&&\left. -\frac{12}{\sqrt{10}}
(E^{(1)}_2 M^{(2)}_2 - M^{(1)}_2 E^{(2)}_2)
[Y_2 (\Omega_1)\otimes Y_2 (\Omega_2)]^1_1 \right)\\
W_{TT}&=&\frac{(2\pi)^3}{4}\left(
-\frac{3}{4\pi}\sum_i \sum_l |E^{(i)}_l|^2 -\frac{3}{2\pi}
\sum_l E^{(1)}_l E^{(2)}_l P_l (\cos \omega_{12})\right.\nonumber\\
&&+\frac{12}{\sqrt{15}}\frac{1}{\sqrt{4\pi}}
(M^{(1)}_0 M^{(2)}_2 Y_{22} (\Omega_{2})
+M^{(1)}_2 M^{(2)}_0 Y_{22} (\Omega_{1}))\nonumber\\
&&+\frac{6\sqrt{7}}{5\sqrt{3}}M^{(1)}_2 M^{(2)}_2
[Y_2 (\Omega_1)\otimes Y_2 (\Omega_2)]^2_2\nonumber\\
&&\left. +\frac{6}{\sqrt{10}}
(E^{(1)}_2 M^{(2)}_2 - M^{(1)}_2 E^{(2)}_2)
[Y_2 (\Omega_1)\otimes Y_2 (\Omega_2)]^1_1 \right)\\
\label{wlt}
W_{LT}&=&\frac{(2\pi)^3}{4}\left(
\frac{3}{4\pi}\sum_i \sum_l E^{(i)}_l C^{(i)}_l \right.\nonumber\\
&&+\frac{3}{4\pi} \sum_l (C^{(1)}_l E^{(2)}_l +E^{(1)}_l C^{(2)}_l )
P_l (\cos \omega_{12})\nonumber\\
&&\left. -\frac{3}{\sqrt{10}}(C^{(1)}_2 M^{(2)}_2 -M^{(1)}_2 C^{(2)}_2)
[Y_2 (\Omega_1)\otimes Y_2 (\Omega_2)]^1_1 \right),
\end{eqnarray}
with $\Omega_i$ the solid angle of $\vec{k}_i$ and $\omega_{12}$
 the angle between $\vec{k}_1$ and $\vec{k}_2$. The quantities
$C,E,M$ are defined as (the upper and lower signs  correspond
to $i=1,2$ respectively):
\begin{eqnarray}
C^{(i)}_l &=&\frac{1}{2m}(\pm 2p\cos\theta -Q)G_E^{(i)}
\phi_l^{(i)}\nonumber\\
E^{(i)}_l &=&\frac{1}{2m}(\mp \sqrt{2}p\sin\theta)G_E^{(i)}
\phi_l^{(i)}\nonumber\\
M^{(i)}_l &=&\frac{1}{2m}Q G_M^{(i)}
\phi_l^{(i)}.
\end{eqnarray}
They are associated with the longitudinal (C) and transverse (E)
part of the convection current, and with the  magnetic (M) current.

The deuteron wave function was normalized as:
\begin{equation}
\sum_{l=0,2} \int dk k^2 |\phi_l (k)|^2 =1.
\end{equation}
The structure functions corresponding to the nuclear current
in eq. (\ref{2bod}) (with $S=1$) can be obtained
from eqs. (\ref{wl}-\ref{wlt})
by omitting the terms with $l=2$.

\section{Applications}
\subsection{Results for the $A=2$ system}
\label{s4.1}
We first compare $e^+ e^-$ production for $pp$ and $pn$ capture
into a bound S-state
(see section \ref{s3}) for 200 MeV incident protons. This corresponds to
virtual photons with invariant mass up to about 100 MeV.

In figure \protect\ref{fig3}
we show the $(p,e^+ e^-)$ cross section as a function of
the invariant mass. The transverse part is sharply peaked at low
invariant masses, due to the phase-space factor $Q/M$ in
eq.\ (\ref{ef1}).
Therefore its angular dependence will be
similar to that of the $(p,\gamma)$ cross section.
 The longitudinal part, on the other hand,
gets its main contributions at intermediate and
high invariant masses, due to the additional factor $M^2 /Q^{2}_0 $
for $W_L$ in eq.\ (\ref{struc}).

We see that the $pp$ cross section
is overall smaller than the $pn$, in both the transverse and
longitudinal part.

The angular dependence of the $(p,e^+ e^- )$ cross section is shown in
figure \protect\ref{fig4}.
Note that the cross section for the $pp$ system is symmetric
around a scattering angle of $90^{\circ}$.
The transverse part has a similar behaviour
as the $(p,\gamma)$ cross section
for both $pp$ and $pn$ systems. One sees that at forward angles
 the
transverse $(p,e^+ e^-)$ cross section for the $pp$ system is much
smaller than that for the $pn$ system (the same holds for the
 $(p,\gamma)$ cross section).
This can be understood from the structure of the nuclear currents
in eq. (\ref{2bod}).
In the case of identical particles 1 and 2, we have $S=0$ for the
final state. As noted in \cite{her}, the magnetic
current then requires $S'=1$ for the initial state, and
destructive interference will occur. The same also holds for the
part of the convection current proportional to $\vec{p}$. Therefore the
transverse part of the cross section is severely
suppressed. It even has a sharp
zero at $90^{\circ}$ in this simple model for the hadron current,
though this feature
will be eliminated by higher order terms in the non-relativistic
expansion of the current operator (such as the relativistic spin
correction \cite{her}), and distortion effects.

 One would  expect the longitudinal part of the
$(p,e^+ e^-)$ cross section to become dominant for $pp$ capture,
because the charge density operator is unaffected by interference
effects. However, the PWIA treatment of the transition charge density
is incorrect at low three-momentum of the photon \cite{ama},
since in that case orthogonality
of initial and final state strongly suppresses the longitudinal response.
In a PWIA approach one should preferably use the longitudinal current
instead of the charge density. By inspecting  eq. (\ref{2bod}) it is
clear that the part of the longitudinal current proportional to
$p(G^{(1)}_E \phi_0^{(1)}-G^{(2)}_E \phi_0^{(2)})$ interferes
destructively, which leads to a further suppression.
As a result the longitudinal cross section is also for the $pp$ system
smaller than the transverse cross section, except in the region
around $90^{\circ}$ where the tranverse response vanishes in the
present model.

This is also reflected in the  conversion ratios
(integrated over invariant mass),
 which are shown in figure \protect\ref{fig5}. The conversion
ratio is expected to be about $\alpha$ because of the second
electromagnetic interaction vertex. The calculated values are indeed
of the order of $\alpha$, but we see some non-trivial structure.
The conversion
ratios may thus be an interesting observable to gain insight in the
electromagnetic response of nuclear systems.
For both the $pn$ and $pp$ system, $R_T$ has
little angular dependence, showing that the transverse $(p,e^+ e^-)$
response follows roughly the real photon response.

 The longitudinal conversion ratio
$R_L$ is smaller than $R_T$ and peaked at forward and backward
scattering angles in case of $pn$. The same holds for the $pp$ system,
 where additionally an enhancement of $R_L$ is observed near
$90^{\circ}$, where the transverse one-body current vanishes in the
present model.

{}From now on the results are based on the full deuteron wave function
 (D-state included) of \cite{mac}.
Figure \protect\ref{fig6}
shows the energy dependence of the conversion factors for the
case of $pn$ capture.
 Naively one would expect, on the basis of diminishing phase space for
the virtual photon,
that both longitudinal and transverse conversion ratios should decrease
with decreasing proton energy. This is true only at small
incident proton energies, however. The longitudinal $R_L$ first
increases and goes (in this PWIA model) to a maximum at T=10 MeV
before decreasing.

Figures \protect\ref{fig7} and \protect\ref{fig8}
contain the four structure functions as a function of
invariant mass, for two angles. Their contributions to the nuclear
current are of comparable magnitude as the direct L and T terms.
In order to investigate the sensitivity to the nuclear structure,
we also plotted
the result for another parametrization of the deuteron wave function
(table 13 in \cite{mac}), which
has a different ratio of S-state component to D-state component
 in the relevant momentum region.
The sensitivity is about the same in all four structure
functions, and drops out in both conversion ratios.

\subsection{Results for $^{12}$C}

For capture to the $^{12}$C groundstate
the spectral function reduces to the squared $1p3/2$
single-particle wave function in momentum space. Since the $(p,\gamma)$
cross section in PWIA
  has (unphysical) zeros, and thus gives rise to infinities in the
conversion ratios,
we replace the pure PWIA momentum distribution by corresponding
DWIA momentum
distributions taken directly from $^{11}\mbox{B}(p,\gamma)$
calculations \cite{jan,hoi} by dividing out the PWIA kinematical factor.
 In this way we can also include
in an approximative way
 the effect of ISI (initial state interactions)
on the conversion ratios, which are shown in figure \protect\ref{fig9}.
The incident proton energy was 98 MeV.
The first two momentum distributions (labeled HF and RPA) were taken
from \cite{jan}. The labels
refer to the treatment of the ISI: in HF the incident proton was
described as a scattering state of the mean field, in RPA
the ISI are treated through an RPA decription of the $^{11}B +p$ system.
The latter was able to give a fair description of the $(p,\gamma)$
cross section. The same holds for the third momentum distribution
(labeled $np-nh$), taken from the continuum shell model calculation in
\cite{hoi}. The behaviour of the conversion factors is quite
similar to the case of $pn$ capture. Sizeable differences
can be noted
between the three approximations, which can be related
to the different slope of the distorted momentum distributions
 at large momenta.

In the literature experimental values of the conversion factor
for $\nuc{11}{B}(p,e^+ e^-)$ have been reported by the Uppsala
group \cite{wies}.
However, the quoted experimental
values
are much larger than any of the DWIA calculations presented.
At present we have no explanation for this. It is unlikely
that this can be attributed  to the time-like nucleon
form factor, which has a negligible  variation in the region of the
invariant mass of the virtual photon that we are considering here
 (less than 3\% on the current matrix elements).

\section {Summary and discussion}

In this paper we have derived the cross section for
dilepton production in radiative capture of protons on nuclei.
As an illustration we have computed the cross section for $A=1$ and
$A=11$
targets in the PWIA.
Analogously to the $(e,e'p)$ reaction, where the exchanged photon is
space-like, we have decomposed the cross section for $(p,e^+ e^- )$
into four
independent structure functions which can be experimentally separated
by varying the kinematic variables.

In order to see whether the virtual photon process contains
new information compared to the $(p,\gamma)$ reaction we have
studied the dependence of the conversion factors (i.e. the ratio
of the virtual to real photon cross sections)  on invariant mass
of the $e^+ e^-$ pair, scattering angle of the photon, and proton
energy.

We found that the longitudinal conversion factor (which represents
a new degree of freedom since it contains a monopole contribution)
 peaks at rather large invariant mass $M$
and is  small compared to the transverse one which is sharply
peaked at small values of $M$ (and is of order $\alpha$).

There appears to be some sensitivity of the conversion factors to
the details of the nuclear structure (contained in the vertex
function) for the
case of $p+\nuc{11}{B}\rightarrow \nuc{12}{C}(g.s.)+e^+ e^-$.
The preliminary data from Uppsala do not agree with our calculations.

A unique feature of the dilepton production is that it allows one to
study the nucleon form factor in the (unphysical) time-like region.
However, for the momentum  transfer range considered here
we found only a small effect (less than 3\%).

In future work we will extent the present PWIA treatment to include
initial state interactions and exchange currents. From the present
investigation it appears that orthogonalization of the plane wave
on the final bound state
represents an important correction when the matrix elements of
the charge operator are considered because of the small value
of the momentum transfer.

Finally we wish to extend the formalism to the more general case
of vir\-tu\-al brems\-strahlung in $p+A$ reactions, which is of interest
to understand the NN contribution to dilepton production
at higher energies.

The authors thank Dr.\ B.\ H\"{o}istad for his interest in this work,
and N.\ Kalantar-Nayestanaki, J.\ Bacelar and H.\ Wilschut for useful
discussions.

\begin{figure}[fig1]
\caption{Kinematical variables for the $A(p,e^+ e^- )A+1$ reaction in
the Impulse Approximation.}
\label{fig1}
\end{figure}
\begin{figure}[fig2]
\caption{Geometry of the $A(p,e^+ e^- )A+1$ reaction in the CM frame.}
\label{fig2}
\end{figure}
\begin{figure}
\caption[fig3]{The $(p,e^+ e^- )$ cross section (see
eq.\ (\protect\ref{ef1}) ),
integrated over
$\Omega$, as a function of the invariant mass $M$ of the
virtual photon. The longitudinal (L) and transverse (T) parts are shown
separately. Full line: $pn$ capture. Dashed line: $pp$ capture.
Both cross sections have been divided by the total (integrated over
$\Omega$) $n(p,\gamma)d$
cross section of eq. (\protect\ref{e14}).
The incident proton energy is 200 MeV.}
\label{fig3}
\end{figure}
\begin{figure}
\caption[fig4]{The $(p,\gamma)$ cross section (labeled $\gamma$)
of eq. (\protect\ref{e14})
and the
$(p,e^+ e^- )$ cross section of eq. (\protect\ref{ef1})
(integrated over
$M$), as a function of the CM scattering angle $\theta$ of the
virtual photon. The longitudinal (L) and transverse (T) parts of the
$(p,e^+ e^- )$ cross section are shown
separately. Full line: $pn$ capture. Dashed line: $pp$ capture.
Both cross sections have been divided by the $n(p,\gamma)d$
cross section at $\theta =0^{\circ}$.
The incident proton energy is 200 MeV.}
\label{fig4}
\end{figure}
\begin{figure}[fig5]
\caption{Longitudinal (L) and transverse (T)
conversion factors $R_i (\theta)=\int dM R_i (\theta,M)$
(see eq.\ (\protect\ref{ef2})),
as a function of the CM scattering angle $\theta$ of the photon.
Full line: $pn$ capture. Dashed line: $pp$ capture. The conversion
factors are divided by the fine-structure constant $\alpha$.
The incident proton energy is 200 MeV.}
\label{fig5}
\end{figure}
\begin{figure}[fig6]
\caption{Longitudinal (L) and transverse (T)
conversion factors $R_i (\theta)=\int dM R_i (\theta,M)$
 for $pn$ capture, for three energies of the incident proton.
The D-state component in the deuteron wave function is taken into
account. Full, short-dashed and long-dashed lines correspond to
incident proton energies of 100,150 and 200 MeV.}
\label{fig6}
\end{figure}
\begin{figure}[fig7]
\caption{The four nuclear structure functions $W_i (\theta,M)$
(see eqs.\ (\protect\ref{wl}-\protect\ref{wlt}))
for the $n(p,e^+ e^-)d$ reaction at an incident proton energy of
200 MeV and CM scattering angle of $\theta =5^{\circ}$, as a
function of the
invariant mass of the virtual photon.
The D-state component in the deuteron wave function is taken into
account. Two parametrizations of the
deuteron wave function are compared.
Full line: Full (energy dependent) model of ref.\ \protect\cite{mac}.
 Dashed line: energy independent model of ref.\ \protect\cite{mac}.}
\label{fig7}
\end{figure}
\begin{figure}[fig8]
\caption{Same as figure 7, for $\theta =45^{\circ}$.}
\label{fig8}
\end{figure}
\begin{figure}[fig9]
\caption{ Longitudinal (L) and transverse (T) conversion factor
$R_i (\theta)=\int dM R_i (\theta,M)$
for the reaction $p+\nuc{11}{B}\rightarrow \nuc{12}{C}(g.s.)
+e^+ +e^-$, at an incident proton energy of 98 MeV. Three theoretical
distorted-wave momentum distributions \protect\cite{jan,hoi} are
compared
(see text):
RPA (full line), HF (short-dashed line) and $np-nh$ (long-dashed line).
The experimental points are taken from \protect\cite{wies}.
They should be compared to the transverse conversion factor only,
because of the limitation to small invariant masses
(up to about 10 MeV) in this experiment.
The conversion
factors are divided by the fine-structure constant $\alpha$.}
\label{fig9}
\end{figure}
\end{document}